\documentclass[fleqn]{article}
\usepackage{espcrc2}

\usepackage{epsfig}

\newcommand{\Mpi}{M_{\pi}}
\newcommand{\Fpi}{F_{\pi}}
\newcommand{\ovr}{\over}
\newcommand{\gaf}{\gamma_5}
\newcommand{\<}{\langle}
\renewcommand{\>}{\rangle}
\newcommand{\pa}{\partial}
\newcommand{\mr}{\mathrm}
\newcommand{\beq}{\begin{equation}}
\newcommand{\eeq}{\end{equation}}
\newcommand{\bea}{\begin{eqnarray}}
\newcommand{\eea}{\end{eqnarray}}

\title{\vspace{-10mm}%
{\normalsize DESY 03-150, SFB/CPP-03-41\hfill{\tt hep-lat/0309169}}\\[-2mm]
{\normalsize September 2003}\\[5mm]
Exploring two non-perturbative definitions of $c_A$%
\thanks{presented by S.\ D\"urr at Lattice 03, Tsukuba, Japan.}%
\thanks{We thank DFG for support in SFB/TR-9.}%
}

\author{
S.\ D\"urr\address[DESY]{DESY, Platanenallee 6, 15738 Zeuthen, Germany},
M.\ Della Morte\addressmark[DESY]%
}

\begin{document}

\begin{abstract}
We present two determinations of the coefficient $c_A$ in quenched QCD, needed
to build the $O(a)$ improved axial current. The first condition used is the
requirement that the PCAC quark mass, as a function of $x_0$, stays flat for a
non-trivial spatial phase for the fermions in the Schr\"odinger functional. The
second condition is that the PCAC relation for the ground-state and the first
excited state at finite L give the same quark mass. Our results confirm
previous findings that in the quenched theory the intrinsic $O(a)$ ambiguity of
$c_A$ gets relevant around $\beta\simeq6.0$.
\vspace{-1pc}
\end{abstract}

\maketitle


\begin{figure*}[!t]
\begin{center}
\epsfig{file=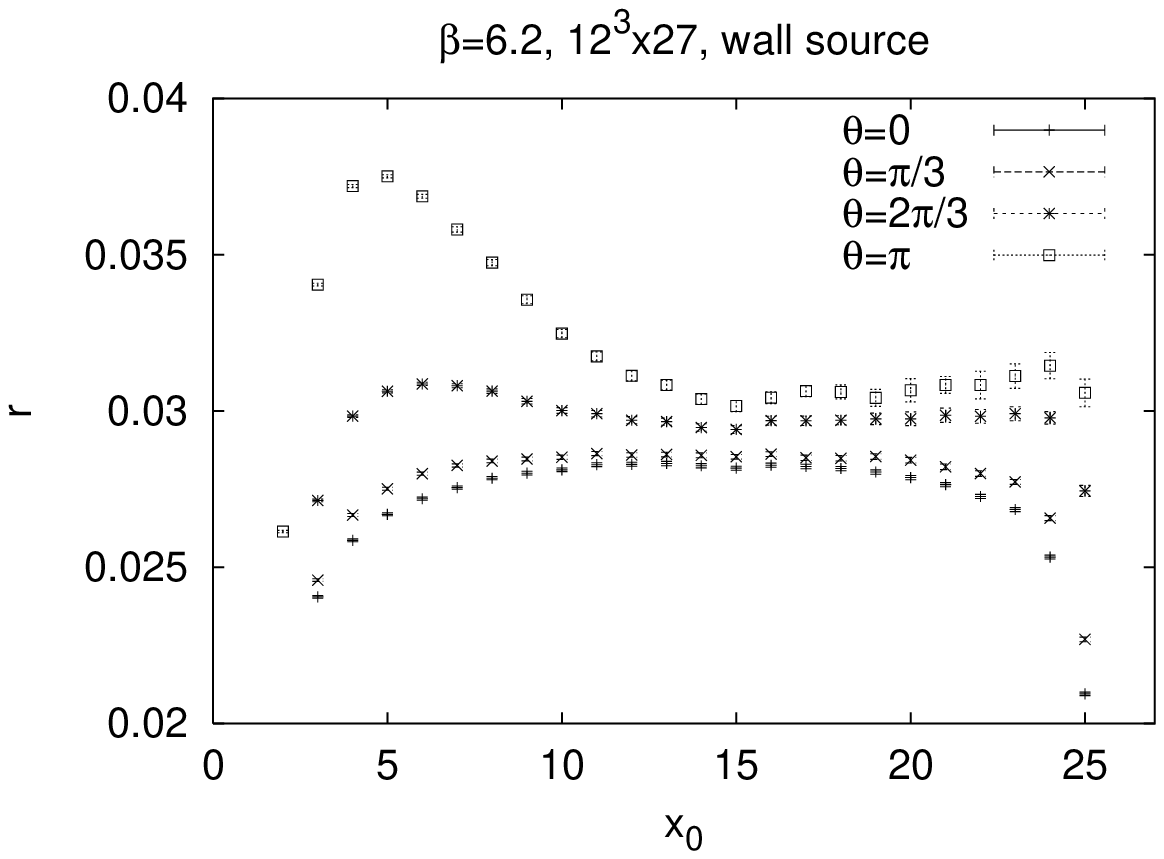,width=75mm,height=50mm}
\epsfig{file=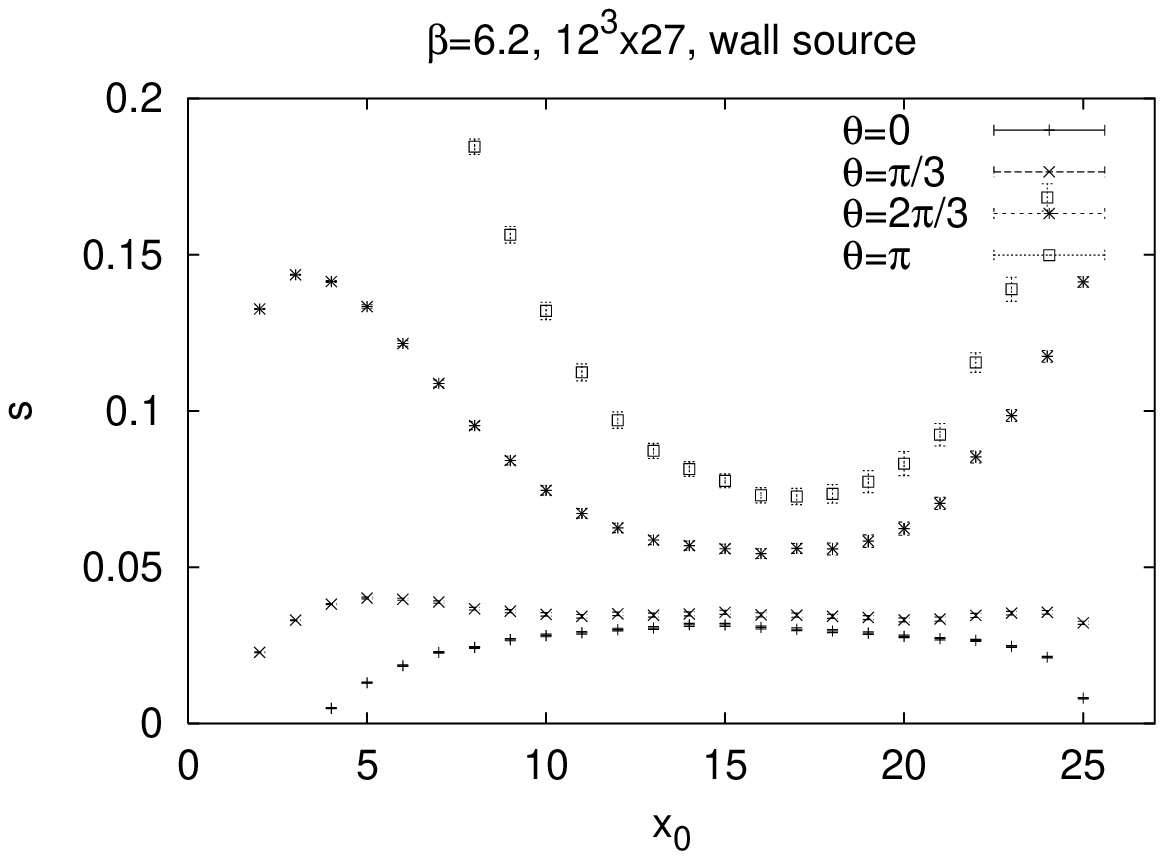,width=75mm,height=50mm}
\end{center}
\vspace{-14mm}
\caption{$r(x_0)$ and $s(x_0)$ from a SF wall source with
four $\theta$-values at $\beta\!=\!6.2, \kappa\!=\!0.13485$.}
\vspace{-2mm}
\end{figure*}


\section{INTRODUCTION}

The improved axial current
\beq
A_0^I(x)\equiv A_0(x)+a\,c_A\,{\pa_0\!+\!\pa_0^*\ovr2}P(x)
\label{a_impr}
\eeq
with $A_0(x)\!=\!\bar\psi_j\gamma_0\gaf\psi_i$ and
$P(x)\!=\!\bar\psi_j\gaf\psi_i$ is designed to modify the scaling behavior of
on-shell quantities from $O(a)$ to $O(a^2)$, if $c_A$ and $c_\mr{SW}$ are
chosen appropriately.
Therefore, an accurate determination of the coefficient $c_A$ is an important
ingredient in the improvement programme \`a la Symanzik.
Here, we present results in the quenched theory, but the real motivation is to
find a criterion which is practical in a dynamical setting, where large
cut-off effects have been found \cite{RainerLat03}.

Our data have been generated in the Schr\"o\-dinger functional (SF) setup: we
use a $T\!\times\!L^3$ box, applying some Dirichlet type boundary conditions in
the time direction (i.e.\ at $x_0\!=\!0, T$) while keeping the gauge-field
periodic in space and the fermions periodic up to a phase:
$U_\mu(x_0,{\bf x}\!+\!L{\bf e}_\mu)=U_\mu(x)$,
$\psi_\mu(x_0,{\bf x}\!+\!L{\bf e}_\mu)=\exp(\mr{i}\theta)\psi_\mu(x)$.
The initial pion with flavor content $ij$ is created at $x_0\!=\!0$ through the
boundary operator
\beq
O_{ij}={a^6\ovr L^3}\sum_{{\bf u}, {\bf v}}
\bar\zeta_i({\bf u})\gaf\zeta_j({\bf v})\;\omega({\bf u}\!-\!{\bf v})
\label{defoij}
\eeq
introducing the wavefunction $\omega$ w.r.t.\ the relative position of the
boundary fields $\bar\zeta_i, \zeta_j$, and absorbed by the local current
$A_0(x)$ in the bulk or by the analogous operator $O_{ij}'$ at $x_0\!=\!T$.
This means that we consider the SF correlators
\bea
f_X(x_0,T,L)\!&\!=\!&\!-{L^3\ovr2}\<X(x)\,O\>
\\
f_1(T,L)\!&\!=\!&\!-\;{1\ovr2}\;\<O'\,O\>
\eea
with $X$ either $A_0$ or $P$ to get the PCAC mass $m\!=\!r\!+\!ac_A\,s$ with
\bea
r(x_0)\!&\!=\!&\!
{{1\ovr2}(\pa_0\!+\!\pa_0^*)f_A(x_0)\ovr2f_P(x_0)}
\label{defr}
\\
s(x_0)\!&\!=\!&\!
{\pa_0\pa_0^*f_P(x_0)\ovr2f_P(x_0)}
\label{defs}
\;.
\eea
Requiring $m$ constant at fixed $\beta$ leads to a definition of the
improvement coefficient through $c_A\!\equiv\!\Delta r/\Delta s$, where the
difference may be w.r.t.
\begin{itemize}
\item[{\bf (0)}]
$\theta$ at fixed $x_0\!=\!T/2\;$ (ALPHA \cite{Luscher:1996ug})
\item[{\bf (1)}]
$x_0$ at fixed $\theta\;$ (LANL \cite{Bhattacharya:1999uq}, ``slope criterion'')
\item[{\bf (2)}]
state in $O[\omega]\,$ (UKQCD \cite{Collins:2001mm}, ``gap criterion'')
\end{itemize}
and depending on the choice, physical quantities differ by $O(a^2)$ effects.
This means that there is an intrinsic $O(a)$ ambiguity in $c_A$ itself
which, already for $\beta\!\simeq\!6$, is not such a small effect
\cite{Bhattacharya:1999uq,Collins:2001mm}.

Here we investigate (in a quenched setting) which one, out of
{\bf (0)}-{\bf (2)}, might be a promising criterion for $N_{\rm f}\!=\!2$,
with a view on the following wishlist:
($i$) no high-energy state involved, i.e.\,no $x_0\!<\!r_0$ (say) used,
($ii$) large ``sensitivity'', i.e.\,not too small value of $\Delta s$,
($iii$) affordable numerical effort, i.e.\,not requiring large volume,
($iv$) ``scalability'', i.e.\,allowing to move to another $\beta$ while keeping
physics in units of $r_0$ constant.
Both the ambiguity of improvement coefficients and how to deal with it have
been discussed in \cite{Guagnelli:2000jw}.

We emphasize that the SF states generated by the boundary operator
(\ref{defoij}) are multiplicatively renormalizable.
The associate $Z$-factor cancels in the ratios (\ref{defr}, \ref{defs}), and
everything is scalable.


\section{THETA CRITERION}

The old ALPHA criterion \cite{Luscher:1996ug} resulted in rather small
$\Delta s$ values.
Furthermore, for $N_\mr{f}\!=\!2$ several $\theta$-angles mean several
simulations.
Therefore, we didn't investigate {\bf (0)} further.


\section{SLOPE CRITERION}

The slope criterion {\bf (1)} requires only one $\theta$-value, but for
completeness we decided to test it for $\theta\!=\!0, \pi/3, 2\pi/3, \pi$.

\begin{figure}[!b]
\vspace{-10mm}
\begin{center}
\epsfig{file=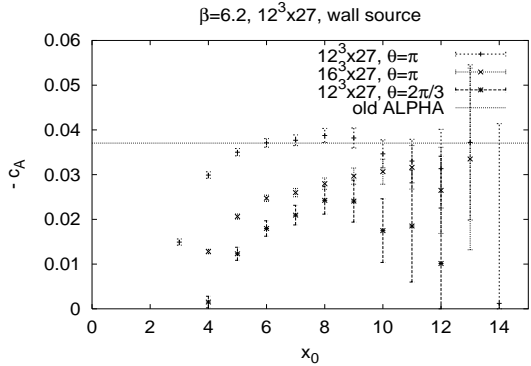,width=75mm,height=50mm}
\end{center}
\vspace{-14mm}
\caption{-$c_A$ via slope in Fig.\,1 with $x_\mr{ref}\!=\!15$.}
\end{figure}

Fig.\,1 shows $r(x_0)$ and $s(x_0)$ in the SF.
For $\theta\!=\!\pi$ there is a good sensitivity.
Using $\Delta r(x_0)\!=\!r(x_0)\!-\!r(x_\mr{ref})$ with $x_\mr{ref}$ around
the extremum (and ditto for $\Delta s$) the recipe $c_A\!=\!\Delta r/\Delta s$
yields a local $c_A(x_0)$.
The remnant dependence on $x_\mr{ref}$ was checked to be small.
This and the dependence on $\theta, L, T, \kappa, \omega$ represent genuine
$O(a)$ effects on $c_A$.

Fig.\,2 displays $-c_A(x_0)$ determined via this ``slope criterion''.
For $\theta\!=\!\pi$ and small enough $L$ there is an early plateau which is a
sign that all states but the lowest two have disappeared.
The corresponding value for $c_A$ seems consistent with the old ALPHA
determination \cite{Luscher:1996ug}.


\section{GAP CRITERION}

\begin{figure*}[!t]
\begin{center}
\epsfig{file=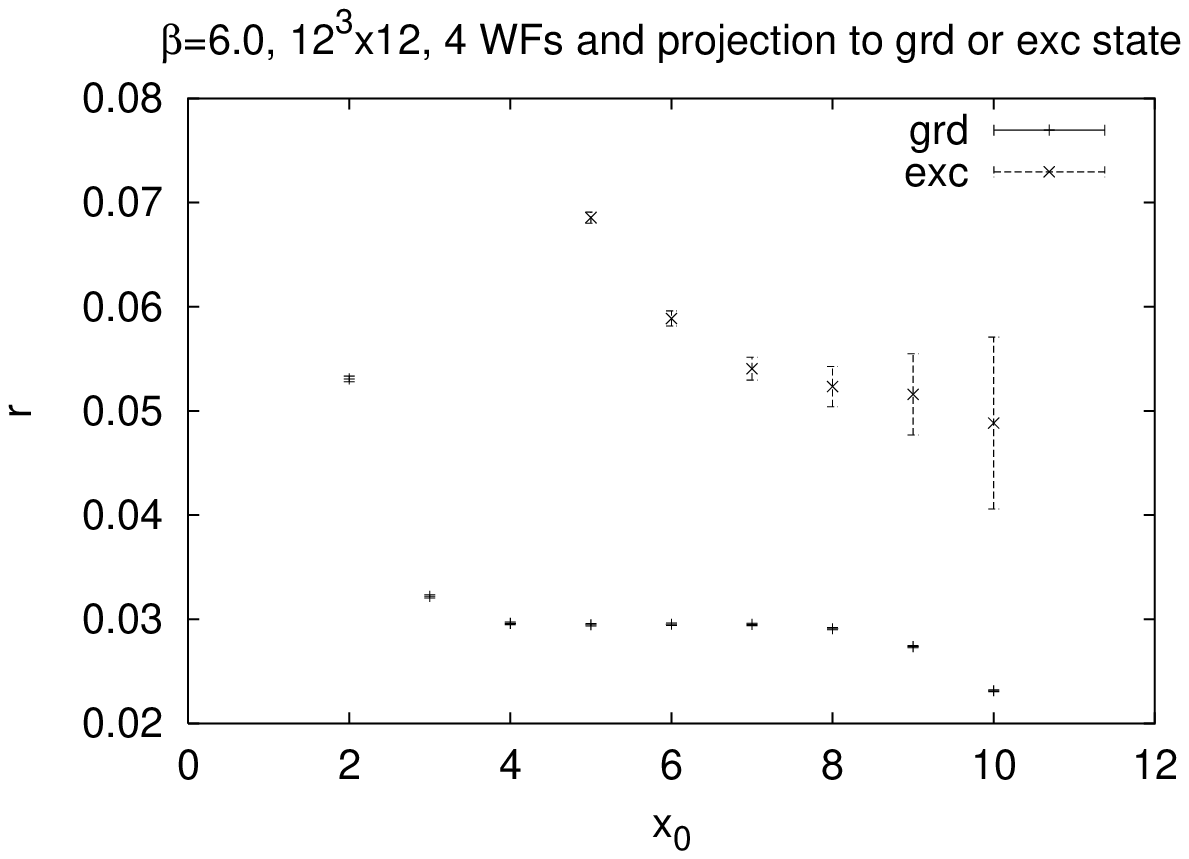,width=75mm,height=50mm}
\epsfig{file=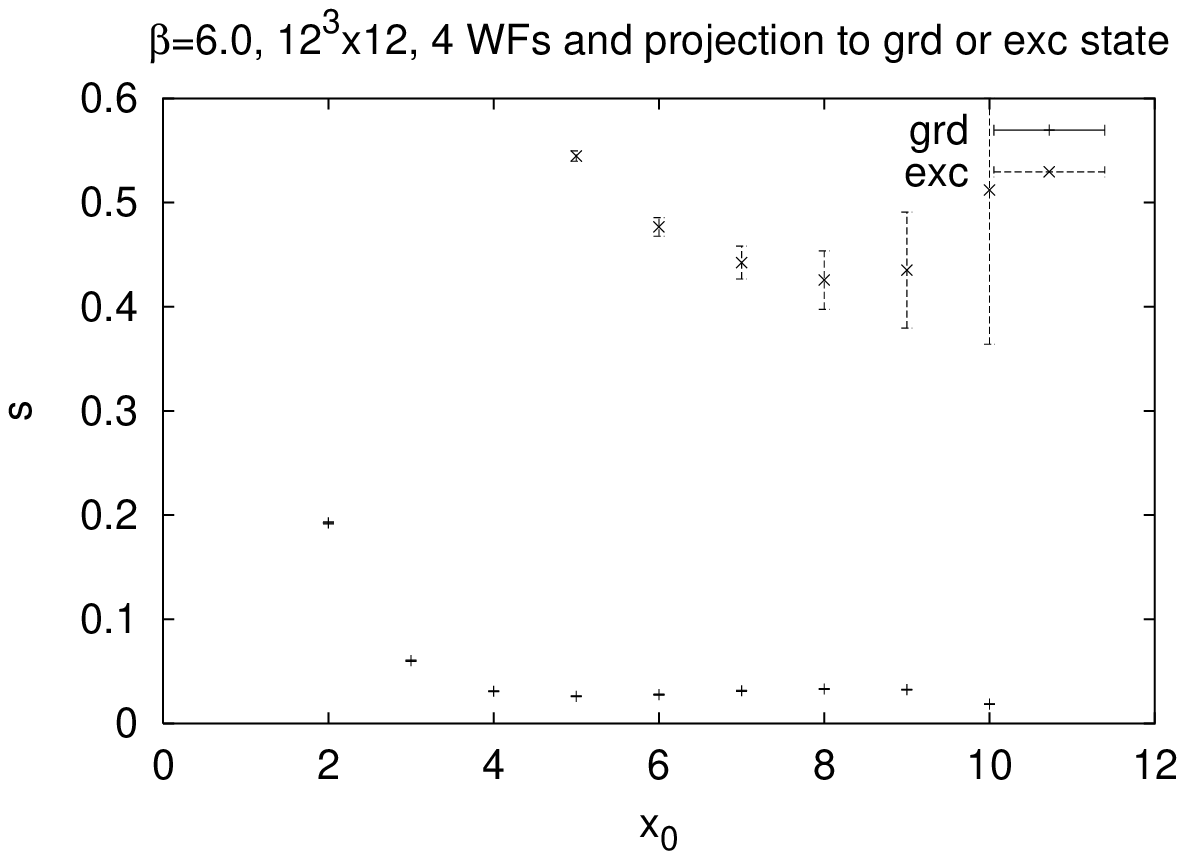,width=75mm,height=50mm}
\end{center}
\vspace{-14mm}
\caption{$r(x_0)$ and $s(x_0)$ from the ground-state and the first excited state
at $\beta\!=\!6.0, \kappa\!=\!0.13415, \theta\!=\!0$.}
\vspace{-2mm}
\end{figure*}

From a transfer matrix analysis one gets
\bea
f_X\!\!\!&\!\!\!\simeq\!\!\!&\!\!\!{L^3\ovr2}\rho\,\xi\,e^{-\Mpi x_0}
\{1\!+\!\eta_X^\pi e^{-\Delta x_0}\!+\!\eta_X^0 e^{-M_G (T\!-\!x_0)}\}
\nonumber
\\
f_1\!\!\!&\!\!\!\simeq\!\!\!&\!\!\!{1\ovr2}\,\rho^2\,e^{-\Mpi T}
\label{origcorr}
\eea
with the matrix element $\xi\!\equiv\!\<0,0|X\!\!\!\!X|\pi,0\>$ and known
representations of $\rho, \eta_X^{\pi, 0}$ in terms of states $|Q,n\>$ with
a given set of quantum numbers and excitation level.
In (\ref{origcorr}) $\Delta$ denotes the gap in the corresponding (here:
pseudoscalar) channel and $M_G$ the mass of the lowest ($0^{++}$) glueball
state.
The important point is that the coefficients $\rho, \eta_X^\pi$ depend on the
initial state $|i_\pi\>$ and hence on the wave function $\omega$, while
$\xi, \eta_X^0$ do not.
This creates the possibility to linearly combine the correlators
(\ref{origcorr}) over several $\omega$ to build one which is clearly dominated
by either the ground-state or the first excited state.
After checking that the corresponding effective masses are indeed distinct
(i.e.\ $M_\pi\!<\!M_\pi^\star$), one may define $\Delta r$ as the difference of
the expressions (\ref{defr}) w.r.t.\ these two linear combinations (at a given
$x_0, \theta$) and ditto for $\Delta s$.
This amounts to an operational definition --~time slice by time-slice~--
of $c_A$ according to the ``gap criterion'' {\bf (2)}.

Fig.\,3 shows $r$ and $s$ at $\theta\!=\!0$ for the ground- and excited state,
built (a posteriori) from the four hydrogen-type wave functions used in the
simulation.
It seems this method yields a rather good sensitivity $\Delta s$.
In the supposed ground-state both $r(x_0)$ and $s(x_0)$ happen to be almost
flat from a rather early time on.

\begin{figure}[!b]
\vspace{-8mm}
\begin{center}
\epsfig{file=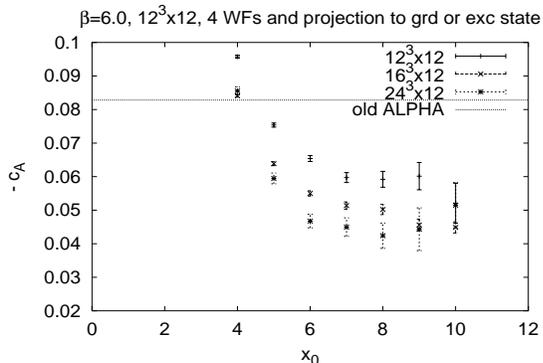,width=75mm,height=50mm}
\end{center}
\vspace{-14mm}
\caption{-$c_A$ via gap in Fig.\,3; ditto at $L\!=\!16,24$.}
\end{figure}

Fig.\,4 shows $-c_A(x_0)$ determined via this ``gap criterion''.
The plateau is reached at $x_0\!\simeq\!0.75\;\mr{fm}$, and typical values are
somewhat smaller than the original ALPHA value at $\beta\!=\!6.0$
\cite{Luscher:1996ug}.
The absolute value $|c_A|\!=\!-c_A$ that we get decreases with $L$, but for
$L\!>\!3r_0$ this effect seems not particularly pronounced any more.
We take this as a sign that there is good hope to determine $c_A$ in a
$(1.5\,\mr{fm})^4$ box in the $N_\mr{f}\!=\!2$ theory.


\section{SUMMARY}

We have tested several improvement conditions that might be used to determine
$c_A$ non-perturbatively in an unquenched setting.
The $O(a)$ ambiguity of $c_A$ that has been pointed out previously
\cite{Bhattacharya:1999uq,Collins:2001mm} has been confirmed via,
$(a)$ extension of {\bf (1)} to several $\theta$ and,
$(b)$ refinement of {\bf (2)} with a wave function projection technique which
achieves dominance by ground- or 1st excited state in the region
0.5\,fm\,...\,1.0\,fm.
We hope this proves sufficient to determine  $c_A$ with $N_\mr{f}\!=\!2$.


\end{document}